\documentstyle[12pt,newpasp,twoside, epsf]{article}

\newcommand{\ie}{{\em i.e.},\ }
\newcommand{\eg}{{\em e.g.},\ }
\newcommand{\etal}{{\em et al.}\ }
\newcommand{\cf}{{\em cf.}\ }
\newcommand{\etc}{{\em etc.}\ }

\def\edcomment#1{\iffalse\marginpar{\raggedright\sl#1\/}\else\relax\fi}
\reversemarginpar

\begin{document}
\title{The Digital Sky Project: Prototyping Virtual Observatory Technologies}
\author{R.J. Brunner}
\affil{Department of Astronomy, California Institute of Technology, Pasadena, CA, 91125}
\author{T. Prince}
\affil{Department of Physics, California Institute of Technology, Pasadena, CA, 91125}
\author{J. Good, T. H. Handley, \& C. Lonsdale}
\affil{IPAC, California Institute of Technology, Pasadena, CA, 91125}
\author{S.G. Djorgovski}
\affil{Department of Astronomy, California Institute of Technology, Pasadena, CA, 91125}

\begin{abstract}

Astronomy is entering a new era as multiple, large area, digital sky
surveys are in production. The resulting datasets are truly remarkable
in their own right; however, a revolutionary step arises in the
aggregation of complimentary multi-wavelength surveys (\ie the
cross-identification of a billion sources). The federation of these
large datasets is already underway, and is producing a major paradigm
shift as Astronomy has suddenly become an immensely data-rich field.
This new paradigm will enable \emph{quantitatively and qualitatively
new science}, from statistical studies of our Galaxy and the
large-scale structure in the universe, to discoveries of rare,
unusual, or even completely new types of astronomical objects and
phenomena. Federating and then exploring these large datasets,
however, is an extremely challenging task. The Digital Sky project was
initiated with this task in mind and is working to develop the
techniques and technologies necessary to solve the problems inherent
in federating these large databases, as well as the mining of the
resultant aggregate data.

\end{abstract}


\section{Introduction}

The Digital Sky project is an NPACI (National Partnership for Advanced
Computing Infrastructure, an NSF Computer Science initiative)
sponsored program to study the role of advanced computational systems
(both processor and network oriented) in the distribution of data from
multiple large area, digital sky surveys. This project was originally
conceived by Tom Prince, the principal investigator for the project,
in 1996 as way to leverage high performance computing to tackle some
of the incipient problems involved in federating and mining the large
amounts of Astronomical information that were beginning to become
available. The project was initially funded in 1997, and has since
grown to include a large number of participants from Caltech
Astronomy, the Caltech Center for Advanced Computing Research, the
Infrared Processing and Analysis Center, the Jet Propulsion
Laboratory, and the San Diego Center for Supercomputing.

One of the fundamental tenets of the Digital Sky project is the
requirement not to develop another analysis tool (which is the last
thing needed by the Astronomical community). Instead, we conceived
that the project would be able to optimally achieve its goals by
serving as a technology demonstrator. Initially, we focused on
identifying the set of requirements to create a prototype virtual
observatory (\ie the ``Digital Sky''). Afterwards, we researched the
core problem of astronomical data federation. Currently, we are
exploring the concept of image data mining.

\begin{figure}[!htb]
\plotfiddle{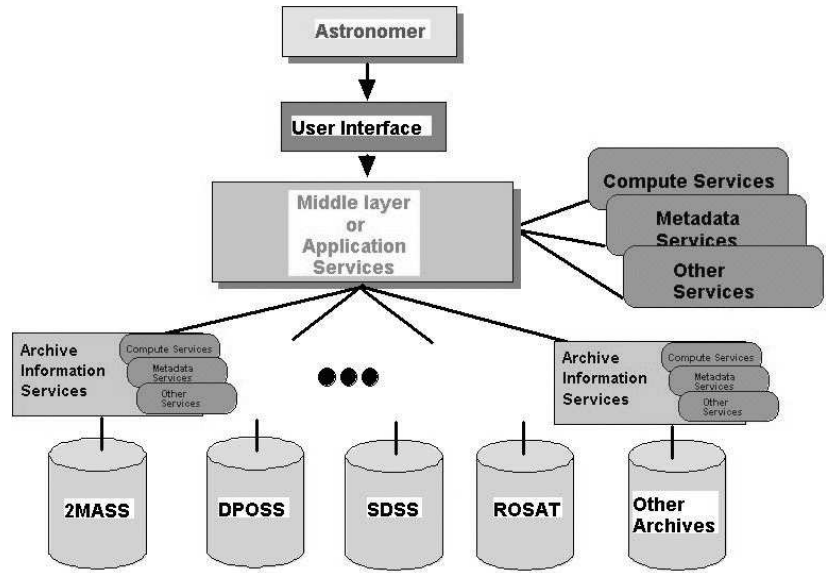}{3.0 in}{0}{40}{40}{-144}{0}
\caption{
\small A top-down overview of the architecture for a virtual observatory. 
The ability to view either a seamless view of the sky or the detailed
capabilities of any individual archive is the overriding
requirement.}
\end{figure}

Overall, our approach is designed to employ a minimalist design, that
focuses more on integrating existing components in cooperation with
community experts (see Figure 1 for an overview). Any knowledge or
applications developed in the course of this project have been
disseminated to the appropriate experts within the community who are
in interested in identifying and attempting to solve the complex
problems that arise in federating disparate, large area, digital sky
surveys.

\section{Towards a Virtual Observatory}

The first major task we tackled was the identification of the types of
issues which needed to be successfully addressed in order to
seamlessly federate highly distributed datasets in order to facilitate
knowledge discovery (see also, Szalay \& Brunner 1998). This highly
desirable end-result is now more commonly known as a virtual
observatory. Formally, we split our list of identified issues into two
categories: basic and advanced, based on the difficulty of
successfully implementing the appropriate service.

The basic services listed below often are provided as part of a
commercial database system, which most of the major archive centers
already utilize.

\begin{description}
\item[Catalog Search Engines.] In order to be fully operable within a virtual 
observatory framework, a dataset must be able to support, at a
minimum, basic query functionality, such as spatial range queries.

\item[System Metadata.] In order to develop general purpose tools, as well as 
simplify the learning process, archives need to implement a
standardized format for describing both the data contained within an
archive and the services which the archive can perform.

\item[Relationship Generators.] A virtual observatory must be able to 
support the cross-identification of billions of sources in both a
static and dynamic state over thousands of square degrees in a
multi-wavelength domain (Radio to X-Ray).

\item[Image Metadata Search Engines.] Image data should be able to be 
selected based on the actual metadata of the images, for example,
spatial location, observational date, \etc

\item[Image Archive Access.] Archived image data should be accessible to an end user,
even if it is merely served from an FTP site.

\item[Query Optimizations.] Within a virtual observatory, a query 
could be distributed to multiple archives. As a result certain
optimizations can be performed depending on the status of the
underlying network topology (\eg network weather) in order to balance
the resulting server load. Optimistically, a learning mechanism can be
applied to analyze queries, and using the accumulated knowledge gained
from past observations (\ie artificial intelligence), queries can be
rearranged in order to provide further performance enhancements.

\end{description}

The following set of services are more advanced, requiring extra
effort, beyond the more common basic services, to be fully
implemented. These services often require post-processing of extracted
data in order to be performed, and can, therefore, make use of
specialized hardware.

\begin{description}

\item[Computational \& Data Grid.] Numerous, complex queries will swamp 
traditional archive topologies. Instead, a virtual observatory should
be implemented to minimize network traffic by utilizing advanced
computational systems to perform complex analysis (\eg correlation
analysis) on the server side as opposed to the client. This solution
can efficiently capitalize on parallel I/O and computing resources, as
well as replicated and persistent datasets, in order to simplify the
implementation of the following advanced services.

\item[Image Processing.] Often, a user will want to post-process an image data request
in order to obtain a scientifically useful result. In order to
implement this feature, basic image processing operations need to be
available on the image server, or close to it, to perform, at a
minimum, basic image operations such as mosaicing, sub-setting and
registration.

\item[Statistical Analysis.] In some cases, a user will be more interested in a
particular statistical analysis of a query result than the actual
resulting dataset itself (\eg a histogram, statistical measure, or
cluster finding code). In such cases, a user of a virtual observatory
should be able to filter the query result using either an available
toolkit or else custom developed statistical codes as necessary.

\item[Visualization.] Undoubtedly, a major component of a virtual observatory is
the ability to visualize data, which might stem from simple graphical
representations of catalog data (as part of a statistical analysis),
seamless serving of image data, or virtual explorations of parameter
space. In certain scenarios, such as defining a new aggregate class of
objects (\eg clusters) or to aid in the mining of the aggregate data,
this process is integrally linked to the actual analysis which is the
desired end-goal.

\item[Machine Learning.] When exploring the forthcoming datasets, especially 
after they have been federated, traditional techniques will quickly be
swamped and rendered hopelessly antiquated. The only efficient
technique to explore the vast, newly opened portions of parameter
space is to capitalize on the inherent capabilities of the very
compute resources which are facilitating the construction of the
virtual observatory. These capabilities give rise to the adoption of
algorithms which let the computer mine for the priceless nuggets in
the mountains of data, and include techniques which can be either
supervised (\eg find everything similar to this particular object) or
unsupervised (\eg find interesting things).

\end{description}

The overriding design principle that we have advocated is to
encapsulate the archival services (both in design and implementation)
which will simplify the effort required to provide interoperability
between different archives (\ie a plug-n-play model). This approach
allows for future growth by providing a blueprint for new archives to
follow and thereby capitalize on existing infrastructures via the
adoption of community standards. This common service approach, reduces
the overall cost to the community by providing a standardized code
base, and facilitates interoperability for analysis tool providers.

In order to be successful, a virtual observatory must be able to grow
through the incorporation of new surveys and datasets in addition to
its original tenants. This requirement necessitates the adoption of
archival standards for exchanging not only the actual data, but also
both metadata and metaservices between constituent archives. This will
allow for different analysis tools to be able to seamlessly work with
data extracted from different archives. All of this work will
culminate in the creation of a National (and eventually Global)
Virtual Observatory, which will eventually enable and empower
scientists and students anywhere to do important, cutting-edge
research.

\section{Relationship Generators}

Before any advanced data exploration or mining tools can be employed,
however, the data of interest must be federated. Indeed, this data
federation service is one of the primary requirements for the National
Virtual Observatory (NVO). Federating these different datasets,
however, is a challenging task.  As this project's initial focus, we
researched solutions to the problems inherent in the dynamic,
multi-wavelength cross-identification of large numbers of Astronomical
sources.

\begin{figure}[!htb]
\plotfiddle{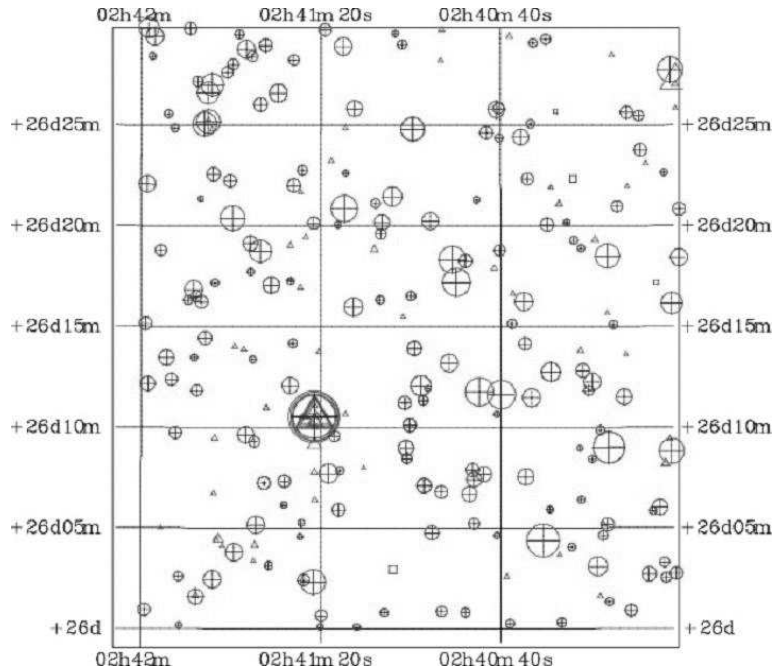}{3.5 in}{0}{40}{40}{-144}{0}
\caption{
\small
A cross-identification example between 2MASS and DPOSS. In order to
provide an uncluttered visual image, both datasets were reduced in
size by imposing a magnitude cut. This image demonstrates many of the
common problems in multiple-wavelength data federation, namely high
source densities, non-detections, and source splitting and
deblending.}
\end{figure}

Specifically, we identified the following topics, which we feel must
be addressed in order to develop a robust cross-identification
service. First, in order to handle the large amounts of dynamic,
multi-wavelength information, including custom user datasets, a data
federation framework must utilize the forthcoming national
computational grid to optimally perform the necessary
calculations. These calculations can be quite complex, particularly in
the case of multi-wavelength data which has varying positional
accuracies (i.e., beam widths) or involves different physical
phenomena. As a result, this framework must be able to incorporate not
only user defined probabilistic associations within the federation
algorithms, but also multiple associations (i.e., binaries and
clusters) and previously published results during the
cross-identification process.  Finally, the newly generated data must
be able to remain persistent so that data mining algorithms can be
applied to the (potentially computationally expensive) result.

As an initial technology demonstration, we developed a custom data
federation service which utilized an optimal data-chunking algorithm
allowing it to easily scale with the size of the resulting datasets
(see Figure 2 for a demonstration). The majority of these tests
focused on data from the 2MASS (2 Micron All Sky Survey, \cf Skrutskie
\etal ) and DPOSS (Digitized Palomar Observatory Sky
Survey, Djorgovski \etal 1998) projects.  As a result of its efficacy,
this software was incorporated by the 2MASS survey into its quality
assurance pipeline as well as the Infrared Science Archive (IRSA) as
the core of its data federation service. We also developed
probabilistic association techniques, building on some of the
pioneering work by Lonsdale \etal (1998) in order to federate the
ROSAT bright source catalog (which has relatively large spatial
uncertainties) with available optical datasets (\eg Rutledge \etal
2000). This work is now being extended to provide an unbiased
quantification of quasars and their environments (Brunner \etal 2001,
in preparation).

\section{Image Mining Operations}

With the success of the ongoing cross-identification work, the project
has now refocused on the challenges of exploring image (or pixel)
parameter space. Ideally, a virtual observatory should be able to
generate seamless views of the universe from any available image
dataset. This process, however, is complicated by various
observational artifacts (see Figure 3 for a demonstration). In order
to simplify the development, we have initially separated the overall
task into the generation of scientifically calibrated datasets from
the generation of visually clean images (which are ideally suited for
public outreach).

\begin{figure}[!htb]
\plotfiddle{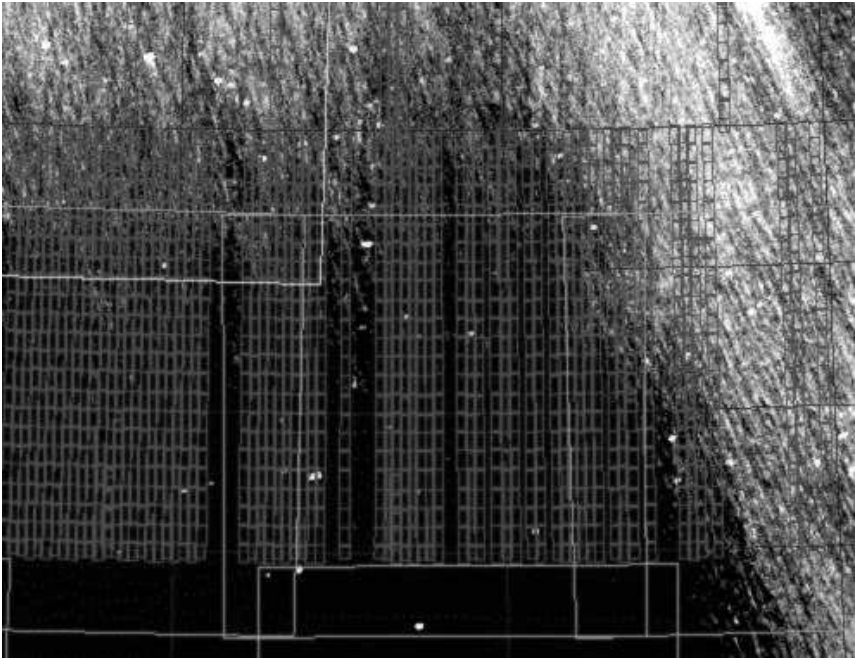}{3.5 in}{0}{65}{65}{-162}{0}
\caption{
\small
Demonstration of some of the difficulties inherent the image
cross-identification project. Superimposed on an IRAS 100 Micron
image, approximately 10 degree square are the outlines for available
DPOSS plates (large squares) and 2MASS images (small rectangles).}
\end{figure}

As a demonstration of the former, we have mosaiced multiple DPOSS
plates in an effort to explore diffuse emission over very large
angular scales (see, \eg Mahabal \etal and Jacob \etal this
volume). The latter project has developed into an astronomical
equivalent of the popular teraserver project, which provides the
ability to pan and zoom around space-based images of Earth. This new
project, aptly named virtualsky, is accesible online at
\mbox{\em http://www.virtualsky.org/} (Williams, R 2000, private
communication), allows a user to pan and zoom around various
astronomical datasets in an analogous fashion.

\acknowledgments

We wish to thank the members of the Digital Sky project, which is
funded through the NPACI program (NSF Cooperative Agreement
ACI-96-19020), and the many Pasadena area virtual observatory
enthusiasts for stimulating discussions.

\end{document}